\documentclass[preprint,amsmath,amssymb,aps,showkeys,showpacs]{revtex4}
\usepackage[english]{babel}
\usepackage{graphicx}
\usepackage{graphics}
\usepackage{amsmath}
\usepackage{dcolumn}
\usepackage{amssymb}
\usepackage{bm}
\usepackage[latin1]{inputenc}

\begin{document}
\title{On the influence of a Rashba-type coupling induced by Lorentz-violating effects on a Landau system for a neutral particle}
\author{K. Bakke}
\email{kbakke@fisica.ufpb.br}
\affiliation{Departamento de F\'isica, Universidade Federal da Para\'iba, Caixa Postal 5008, 58051-970, Jo\~ao Pessoa, PB, Brazil.}

\author{H. Belich} 
\affiliation{Departamento de F\'{\i}sica e Qu\'{\i}mica, Universidade Federal do Esp\'{\i}rito Santo, Av. Fernando Ferrari, 514, Goiabeiras, 29060-900, Vit\'{o}ria, ES, Brazil.}
\email{belichjr@gmail.com}

\begin{abstract}
We study a possible scenario of the Lorentz symmetry violation background that allows us to build an analogue of the Landau system for a nonrelativistic Dirac neutral particle interacting with a field configuration of crossed electric and magnetic fields. We also discuss the arising of analogues of the Rashba coupling, the Zeeman term and the Darwin term from the Lorentz symmetry breaking effects, and the influence of these terms on the analogue of the Landau system confined to a two-dimensional quantum ring. Finally, we show that this analogy with the Landau system confined to a two-dimensional quantum ring allows us to establish an upper bound for the Lorentz symmetry breaking parameters. 
\end{abstract}

\maketitle

\section{Introduction}

The Lorentz symmetry principle, with the development of quantum mechanics, provided a guide for the formulation of a theory which describes the behaviour of elementary particles: the Standard Model. The search for a fundamental theory has brought the notion of symmetry, phase transition and spontaneous symmetry breaking. The mechanism suggested by Higgs {\it et al} consolidated these concepts in the scope of the Weinberg-Salam-Glashow model. Despite of the tremendous success of the Standard Model, it fails to incorporate gravity as a fundamental interaction, therefore, the description of neutrinos consists in massless particles. In view of these limitations, there are some proposals for extensions of the Standard Model. In this work, we focus our study on a line that extends the spontaneous breaking of symmetry by tensorial fields.

The study of the symmetry breaking for relativistic systems can be extended by considering a background given by the spacetime indices of a tensor with rank $n\geq1$. In magnetic systems, it is well-known that the spontaneous breaking of symmetry is performed by a vector background, where the symmetry group $SO\left(3\right) $ is spontaneously broken. A natural extension of this violating background is thinking in a four-vector or tensor background. The background field, in this situation, breaks the symmetry $SO\left(1,3\right)$ instead of the symmetry $SO\left(3\right)$. This line of research is known in the literature as the spontaneous violation of the Lorentz symmetry \cite{extra3,extra1,extra2}. This new possibility of spontaneous violation was first suggested in 1989 in a work of Kostelecky and Samuel \cite{extra3} indicating that, in the string field theory, the spontaneous violation of symmetry by a scalar field could be extended. This extension has an immediate consequence: a spontaneous breaking of the Lorentz symmetry. In the electroweak theory, a scalar field acquires a nonzero vacuum expectation value which yields mass to gauge bosons (Higgs Mechanism). Similarly, in the string field theory, this scalar field can be extended to a tensor field. Nowadays, these theories are encompassed in the framework of the Standard Model Extension (SME) \cite{col} as a possible extension of the minimal Standard Model of the fundamental interactions. For instance, the violation of the Lorentz symmetry is implemented in the fermion section of the Standard Model Extension by two CPT-odd terms: $a_{\mu}\overline{\psi}\gamma^{\mu}\psi$ and $b_{\mu}\overline{\psi}\gamma_{5}\gamma^{\mu}\psi$, where $a_{\mu}$ and $b_{\mu}$ correspond to the Lorentz-violating vector backgrounds. From these fixed vector field backgrounds, Lorentz symmetry breaking effects have been investigated in quantum Hall effect \cite{lin3}, self-adjoint extension \cite{ea,ea2,ea3}, bound states solutions \cite{belich2,bb2,bb3} and geometric quantum phases \cite{belich,belich1,belich3,bbs2,brito,bb,bb4}.

Our interest in this work is to study the nonrelativistic quantum dynamics of a Dirac neutral particle interacting with a field configuration of crossed electric and magnetic fields which stems from Lorentz symmetry breaking effects. We discuss the arising of analogues of the Rashba coupling, the Zeeman term and the Darwin term from the Lorentz symmetry breaking effects. Moreover, we study a possible scenario of the Lorentz symmetry violation background that allows us to obtain an analogue of the Landau system, and discuss the influence of the analogues of the Zeeman energy and the Rashba effect \cite{rasb,rasb2,rasel,rasel2,rasel3,rasel4,rasel5,rasmag} on this Landau system confined to a two-dimensional quantum ring \cite{tan}. Finally, we show that this analogy with the Landau system confined to a two-dimensional quantum ring \cite{tan} allows us to establish an upper bound for the Lorentz symmetry breaking parameters.

The structure of this paper is: in section II, we discuss the nonrelativistic limit of the Dirac equation by applying the Foldy-Wouthuysen approximation \cite{fw,greiner}, and show the arising of a Rashba-like coupling, a Zeeman-like term and a Darwin-like term induced by Lorentz symmetry breaking effects; in section III, we study a possible scenario of the Lorentz symmetry violation background that allows us to build an analogue of the Landau system for a nonrelativistic Dirac neutral particle interacting with a field configuration of crossed electric and magnetic fields, and the influence of the analogues of the Zeeman energy and the Rashba effect on this Landau system confined to a two-dimensional quantum ring \cite{tan}; in section IV, we present our conclusions.

\section{Rashba-like coupling, Zeeman-like term and Darwin-like term induced by a Lorentz symmetry violation background}

In this section, we discuss the nonrelativistic quantum dynamics of a Dirac neutral particle interacting with a field configuration of crossed electric and magnetic fields induced by Lorentz symmetry breaking effects. Furthermore, we show that Lorentz symmetry breaking effects can induce a Rashba-like coupling, a Zeeman-like term and a Darwin-like term in the Schr\"odinger-Pauli equation for a spin-$1/2$ neutral particle.

Recently, we have introduced a new term into the Dirac equation in order to describe the relativistic and nonrelativistic quantum dynamics of a Dirac neutral particle interacting with a field configuration of crossed electric and magnetic fields induced by Lorentz symmetry breaking effects \cite{bb}. We have shown that the wave function of the Dirac neutral particle acquires a geometric quantum phase that stems from the interaction between a fixed time-like 4-vector and a configuration of crossed electric and magnetic fields given by $\phi=-\nu\oint\left(\vec{E}\times\vec{B}\right)$, where $\nu$ is a parameter associated with the Lorentz symmetry breaking, $\vec{E}$ and $\vec{B}$ are the electric and magnetic fields, respectively.

In this work, we intend to obtain a Rashba-like coupling, a Zeeman-like term and a Darwin-like term in the Schr\"odinger-Pauli equation which stems from the interaction between a fixed time-like 4-vector and a configuration of crossed electric and magnetic fields by applying the Foldy-Wouthuysen approximation \cite{fw,greiner}. Therefore, we start by modifying the nonminimal coupling proposed in Ref. \cite{bb} and writing it in the following form:
\begin{eqnarray}
i\gamma^{\mu}\partial_{\mu}\rightarrow i\gamma^{\mu}\partial_{\mu}-\frac{g}{2}\,\eta^{\alpha\beta}\,\bar{F}_{\mu\alpha}\left(x\right)\,\bar{F}_{\beta\nu}\left(x\right)\,\gamma^{\mu}\,b_{\lambda}\,\gamma^{\lambda}\,\gamma^{\nu},
\label{4.1}
\end{eqnarray}
where $g$ is a constant, $b^{\mu}$ corresponds to a fixed $4$-vector that acts on a vector field breaking the Lorentz symmetry violation, the tensor $\bar{F}_{\mu\nu}\left(x\right)$ corresponds to the dual electromagnetic tensor, whose components are $\bar{F}_{0i}=-\bar{F}_{i0}=B_{i}$, and $\bar{F}_{ij}=-\bar{F}_{ji}=\epsilon_{ijk}E^{k}$. The $\gamma^{\mu}$ matrices are defined in the Minkowski spacetime in the form \cite{greiner}:
\begin{eqnarray}
\gamma^{0}=\hat{\beta}=\left(
\begin{array}{cc}
1 & 0 \\
0 & -1 \\
\end{array}\right);\,\,\,\,\,\,
\gamma^{i}=\hat{\beta}\,\hat{\alpha}^{i}=\left(
\begin{array}{cc}
 0 & \sigma^{i} \\
-\sigma^{i} & 0 \\
\end{array}\right);\,\,\,\,\,\,\Sigma^{i}=\left(
\begin{array}{cc}
\sigma^{i} & 0 \\
0 & \sigma^{i} \\	
\end{array}\right),
\label{4.2}
\end{eqnarray}
with $\vec{\Sigma}$ being the spin vector. The matrices $\sigma^{i}$ are the Pauli matrices, and satisfy the relation $\left(\sigma^{i}\,\sigma^{j}+\sigma^{j}\,\sigma^{i}\right)=2\eta^{ij}$. The reason for modifying the nonminimal coupling proposed in Ref. \cite{bb} is that the Lorentz symmetry breaking effects described by the nonminimal coupling (\ref{4.1}) do not induce an effective mass for the Dirac neutral particle given by $M=m+\nu\,B^{2}$, which simplifies our discussion.

Thereby, let us consider a Lorentz symmetry violation background given by a time-like vector $b_{\lambda}=\left(b_{0},0,0,0\right)$, then, the Dirac equation in the presence of the nonminimal coupling (\ref{4.1}) (in Cartesian coordinates) becomes
\begin{eqnarray}
i\frac{\partial\psi}{\partial t}=m\hat{\beta}\psi+\vec{\alpha}\cdot\vec{p}\,\psi+g\,b_{0}\vec{\alpha}\cdot\left(\vec{E}\times\vec{B}\right)\psi-gb_{0}E^{2}\,\psi.
\label{4.3}
\end{eqnarray}

From now on, we apply the Foldy-Wouthuysen approximation \cite{fw,greiner} up to the terms of order $m^{-2}$ in order to obtain the nonrelativistic limit of the Dirac equation (\ref{4.3}). In this approach, we need first to write the Dirac equation in the form:
\begin{eqnarray}
i\,\frac{\partial\psi}{\partial t}=\hat{H}\,\psi,
\label{4.3a}
\end{eqnarray} 
where the Hamiltonian operator $\hat{H}$ of the system must be written in terms of even operator $\hat{\epsilon}$ and odd operators $\hat{\mathcal{O}}$ as: $\hat{H}=m\,\hat{\beta}+\hat{\mathcal{E}}+\hat{\mathcal{O}}$. Both even and odd operators must be Hermitian operators and satisfy the following relations: $\left[\hat{\mathcal{E}},\hat{\beta}\right]=\hat{\mathcal{E}}\,\hat{\beta}-\hat{\beta}\,\hat{\mathcal{E}}=0$ and $\left\{\hat{\mathcal{O}},\hat{\beta}\right\}=\hat{\mathcal{O}}\,\hat{\beta}+\hat{\beta}\,\hat{\mathcal{O}}=0$. In short, the objective of the Foldy-Wouthuysen approach \cite{fw,greiner} is to apply a unitary transformation in order to remove the operators from the Dirac equation that couple the ``large'' to the ``small'' components of the Dirac spinors. We have that the even operators $\hat{\mathcal{E}}$ do not couple the ``large'' to the ``small'' components of the Dirac spinors, while the odd operators $\hat{\mathcal{O}}$ do couple them. Thereby, by applying he Foldy-Wouthuysen approximation \cite{fw,greiner} up to the terms of order $m^{-2}$, we can write the nonrelativistic limit of the Dirac equation in the form:
\begin{eqnarray}
i\frac{\partial\psi}{\partial t}=m\hat{\beta}\psi+\hat{\mathcal{E}}\psi+\frac{\hat{\beta}}{2m}\,\hat{\mathcal{O}}^{2}\psi-\frac{1}{8m^{2}}\left[\hat{\mathcal{O}},\left[\hat{\mathcal{O}},\hat{\mathcal{E}}\right]\right]\psi.
\label{4.4}
\end{eqnarray}

Hence, from Eq. (\ref{4.3}), the operators $\hat{\mathcal{O}}$ and $\hat{\mathcal{E}}$ are
\begin{eqnarray}
\hat{\mathcal{O}}=\vec{\alpha}\cdot\vec{p}+g\,b_{0}\,\vec{\alpha}\cdot\left(\vec{E}\times\vec{B}\right);\,\,\,\,\,\,\,\,\,\hat{\mathcal{E}}=-g\,b_{0}\,E^{2}.
\label{4.6}
\end{eqnarray}

By substituting (\ref{4.6}) into (\ref{4.4}), we obtain the Schr\"odinger-Pauli equation
\begin{eqnarray}
i\frac{\partial\psi}{\partial t}&=&m\hat{\beta}\psi+\frac{\hat{\beta}}{2m}\left[\vec{p}+gb_{0}\left(\vec{E}\times\vec{B}\right)\right]^{2}\psi+\frac{\hat{\beta}}{2m}\,gb_{0}\,\vec{\Sigma}\cdot\vec{B}_{\mathrm{eff}}\,\psi-gb_{0}\,E^{2}\,\psi\nonumber\\
[-2mm]\label{4.7}\\[-2mm]
&+&\frac{gb_{0}}{8m^{2}}\left(\vec{\nabla}\cdot\vec{E}_{\mathrm{eff}}\right)\psi-\frac{gb_{0}}{4m^{2}}\,\vec{E}_{\mathrm{eff}}\cdot\left(\vec{\Sigma}\times\vec{p}\right)\,\psi+\frac{igb_{0}}{8m^{2}}\,\vec{\Sigma}\cdot\left(\vec{\nabla}\times\vec{E}_{\mathrm{eff}}\right)\,\psi+\tilde{\mathcal{O}}\left(\frac{g^{2}}{m^{2}}\right).\nonumber
\end{eqnarray}

Note that we have defined in Eq. (\ref{4.7}) a connection 1-form $A_{\mu}^{\mathrm{eff}}\left(x\right)$ in such a way that the time-like component $A_{0}^{\mathrm{eff}}\left(x\right)$ corresponds to an effective scalar potential, and $\vec{A}_{\mathrm{eff}}\left(x\right)$ corresponds to an effective potential vector, that is,
\begin{eqnarray}
A_{0}^{\mathrm{eff}}\left(x\right)=E^{2};\,\,\,\,\vec{A}_{\mathrm{eff}}\left(x\right)=\vec{E}\times\vec{B}.
\label{4.8}
\end{eqnarray}
In this way, we can define an effective magnetic field and an effective electric field a
\begin{eqnarray}
\vec{B}_{\mathrm{eff}}&=&\vec{\nabla}\times\vec{A}_{\mathrm{eff}}=\vec{\nabla}\times\left(\vec{E}\times\vec{B}\right)\nonumber\\
[-2mm]\label{4.9}\\[-2mm]
\vec{E}_{\mathrm{eff}}&=&-\vec{\nabla}A_{0}^{\mathrm{eff}}=-\vec{\nabla}E^{2}.\nonumber
\end{eqnarray} 

Moreover, we have obtained in Eq. (\ref{4.7}) a Zeeman-Like term $\hat{H}_{\mathrm{Z}}$, a Darwin-like term \cite{greiner} $\hat{H}_{\mathrm{D}}$, and a Rashba-like term $\hat{H}_{\mathrm{R}}$ given by
\begin{eqnarray}
\hat{H}_{\mathrm{R}}&=&-\frac{g\,b_{0}}{4m^{2}}\,\vec{E}_{\mathrm{eff}}\cdot\left(\vec{\Sigma}\times\vec{p}\right);\nonumber\\
\hat{H}_{\mathrm{Z}}&=&\frac{\hat{\beta}}{2m}\,gb_{0}\,\vec{\Sigma}\cdot\vec{B}_{\mathrm{eff}};\label{4.10}\\
\hat{H}_{\mathrm{D}}&=&\frac{g\,b_{0}}{8m^{2}}\left(\vec{\nabla}\cdot\vec{E}_{\mathrm{eff}}\right).\nonumber
\end{eqnarray}

Observe that Eq. (\ref{4.7}) recovers the nonrelativistic limit of the Dirac equation obtained in Ref. \cite{bb} if we consider just terms of order $m^{-1}$ of the Foldy-Wouthuysen approach. By considering the field configuration of Ref. \cite{bb} given by $\vec{E}=\frac{\lambda}{\rho}\,\hat{\rho}$ and $\vec{B}=B_{0}\,\hat{z}$ (where $\rho=\sqrt{x^{2}+y^{2}}$, $\lambda$ is a linear density of electric charges, $B_{0}$ is a constant and $\hat{\rho}$ and $\hat{z}$ are unit vectors in the radial and $z$ directions, respectively), we can see that the presence of the effective potential vector given in Eq. (\ref{4.8}) yields the arising of a geometric quantum phase in the wave function of the spin-$1/2$ neutral particle given by $\phi=-gb_{0}\oint\left(\vec{E}\times\vec{B}\right)\cdot\,d\vec{r}=2\pi\,gb_{0}\,\lambda\,B_{0}$ as in Ref. \cite{bb}. The difference between Eq. (\ref{4.7}) and the nonrelativistic equation obtained in Ref. \cite{bb} is that there is no presence of an effective mass $M=m+\nu\,B^{2}$ in Eq. (\ref{4.7}), which does not invalidate our approach since the term $\nu\,B^{2}$ can be considered very small compared to $m$, thus, it can be neglected.

Hence, by applying the Foldy-Wouthuysen approximation \cite{fw,greiner} up to terms of order $m^{-2}$, we have new possible scenarios of studying Lorentz symmetry breaking effects that can be determined by effective scalar and vector potentials (\ref{4.8}) and by relativistic correction terms that gives rise to a Rashba-like coupling, a Darwin-like term and a Zeeman-like term (\ref{4.10}). Recently, the influence of a Rashba coupling induced by the effects of the Lorentz symmetry breaking (defined by fixed space-like vector and a radial electric field) on a two-dimensional quantum ring has been investigated in Ref. \cite{bb5}. 

In particular, the Rashba coupling is important because it is the first relativistic correction that stems from the spin-orbit interaction. It is responsible for splitting states with the same orbital angular momentum $l$, but with different spin $s$ \cite{greiner,rasb2}. This relativistic correction has been widely studied in recent decades in mesoscopic systems \cite{rasb,rasb2,rasel,rasel2,rasel3,rasel4,rasel5,rasmag}.

\section{Influence of the Rashba-like term on the analogue of the Landau system confined to a two-dimensional ring}

In this section, we discuss the arising of the Landau system and a Rashba-like coupling induced by a Lorentz symmetry violation scenario in the nonrelativistic quantum dynamics for a spin-$1/2$ neutral particle. We show that the quantum dynamics established by the introduction of the nonminimal coupling (\ref{4.1}) can yield an analogue of the Landau system for a neutral particle proposed in Ref. \cite{lin2}. The Landau system established in Ref. \cite{lin2} is based on a moving neutral particle which acquires an electric dipole moment induced by a configuration of crossed electric and magnetic fields, where the Landau system is achieved if the field configuration provides the presence of a uniform effective magnetic field given by $\vec{B}_{\mathrm{eff}}=\vec{\nabla}\times\left(\vec{E}\times\vec{B}\right)$. As we have shown in Eq. (\ref{4.9}), the same effective magnetic field can be yielded by Lorentz symmetry breaking effects. Thereby, in order to obtain the Landau system and a Rashba-like coupling from Lorentz symmetry breaking effects, let us consider the field configuration of Ref. \cite{lin2}: 
\begin{eqnarray}
\vec{E}=\frac{\lambda\,\rho}{2}\,\hat{\rho};\,\,\,\,\,\,\,\vec{B}=B_{0}\,\hat{z},
\label{5.0}
\end{eqnarray}
where $\lambda$ and $B_{0}$ are constants, and $\hat{\rho}$ and $\hat{z}$ are units vector on the radial direction and $z$ direction, respectively. Observe that, with the Lorentz symmetry violation background given by a time-like vector $b_{\lambda}=\left(b_{0},0,0,0\right)$ and the field configuration (\ref{5.0}), we have in (\ref{4.9}) that
\begin{eqnarray}
\vec{B}_{\mathrm{eff}}=-\lambda\,B_{0}\,\hat{z};\,\,\,\,\vec{E}_{\mathrm{eff}}=-\frac{\lambda^{2}\rho}{2}\,\hat{\rho}.
\label{5.0a}
\end{eqnarray}

Note that $\vec{B}_{\mathrm{eff}}\neq\vec{B}$ and $\vec{E}_{\mathrm{eff}}\neq\vec{E}$. From the point of view of Ref. \cite{lin2}, the presence of a uniform effective magnetic field $\vec{B}_{\mathrm{eff}}$ given in Eq. (\ref{5.0a}) gives rise to an analogue of the Landau system for spin-$1/2$ neutral particle with an induced electric dipole moment. By observing that $\vec{E}_{\mathrm{eff}}$ is present only in the terms of order $m^{-2}$ in Eq. (\ref{4.7}), then, if we neglect these terms, we obtain an analogue system of the Landau quantization as established in Ref. \cite{lin2}. However, an interesting case occurs when we consider terms of order $m^{-2}$ in Eq. (\ref{4.7}) because the presence of the effective electric field $\vec{E}_{\mathrm{eff}}$ given in Eq. (\ref{5.0a}) does not break the analogue system of the Landau quantization, but gives rise to a Rashba-like coupling.

Furthermore, we can see the Lorentz symmetry violation background defined by a time-like vector $b_{\lambda}=\left(b_{0},0,0,0\right)$ and the field configuration (\ref{5.0}) possesses a cylindrical symmetry. In what follows, we show the way of we need to write the covariant form of the Dirac equation in order to work with curvilinear coordinates. The mathematical formulation used to write the Dirac equation in curvilinear coordinate is the same of spinors in curved spacetime \cite{weinberg,bd,schu}. As an example, in cylindrical coordinates, the line element of the Minkowski spacetime is writing in the form: $ds^{2}=-dt^{2}+d\rho^{2}+\rho^{2}d\varphi^{2}+dz^{2}$. By applying, thus, a coordinate transformation $\frac{\partial}{\partial x^{\mu}}=\frac{\partial \bar{x}^{\nu}}{\partial x^{\mu}}\,\frac{\partial}{\partial\bar{x}^{\nu}}$ and a unitary transformation on the wave function $\psi\left(x\right)=U\,\psi'\left(\bar{x}\right)$, the Dirac equation can be written in any orthogonal system in the following form \cite{schu}:
\begin{eqnarray}
i\,\gamma^{\mu}\,D_{\mu}\,\psi+\frac{i}{2}\,\sum_{k=1}^{3}\,\gamma^{k}\,\left[D_{k}\,\ln\left(\frac{h_{1}\,h_{2}\,h_{3}}{h_{k}}\right)\right]\psi=m\psi,
\label{5.1}
\end{eqnarray}
where $D_{\mu}=\frac{1}{h_{\mu}}\,\partial_{\mu}$ is the derivative of the corresponding coordinate system and the parameter $h_{k}$ corresponds to the scale factors of this coordinate system. In our case (cylindrical coordinates), the scale factors are $h_{0}=1$, $h_{1}=1$, $h_{2}=\rho$ and $h_{3}=1$. Therefore, the second term in (\ref{5.1}) corresponds to a term called the spinorial connection \cite{schu,b4,bbs2,bd,weinberg}. By introducing the nonminimal coupling (\ref{4.1}), the Dirac equation (\ref{5.1}) becomes
\begin{eqnarray}
i\,\gamma^{\mu}\,D_{\mu}\,\psi+\frac{i}{2}\,\sum_{k=1}^{3}\,\gamma^{k}\,\left[D_{k}\,\ln\left(\frac{h_{1}\,h_{2}\,h_{3}}{h_{k}}\right)\right]\psi-\frac{g}{2}\,\eta^{\alpha\beta}\,\bar{F}_{\mu\alpha}\left(x\right)\,\bar{F}_{\beta\nu}\left(x\right)\,\gamma^{\mu}\,b_{\lambda}\,\gamma^{\lambda}\,\gamma^{\nu}=m\psi.
\label{5.2}
\end{eqnarray}

By following the same steps from Eq. (\ref{4.4}) to (\ref{4.8}) to obtain the nonrelativistic limit of the Dirac equation (\ref{5.1}), then, the Schr\"odinger-Pauli equation becomes
\begin{eqnarray}
i\frac{\partial\psi}{\partial t}&=&\frac{1}{2m}\left[\vec{\pi}+gb_{0}\left(\vec{E}\times\vec{B}\right)\right]^{2}\psi+\frac{gb_{0}}{2m}\,\vec{\sigma}\cdot\vec{B}_{\mathrm{eff}}\,\psi-gb_{0}\,E^{2}\psi+\frac{gb_{0}}{8m^{2}}\left(\vec{\nabla}\cdot\vec{E}_{\mathrm{eff}}\right)\psi\nonumber\\
[-2mm]\label{5.3}\\[-2mm]
&-&\frac{gb_{0}}{4m^{2}}\,\vec{E}_{\mathrm{eff}}\cdot\left(\vec{\sigma}\times\vec{\pi}\right)\,\psi+\frac{igb_{0}}{8m^{2}}\,\vec{\sigma}\cdot\left(\vec{\nabla}\times\vec{E}_{\mathrm{eff}}\right)\,\psi+\tilde{\mathcal{O}}\left(\frac{g^{2}}{m^{2}}\right).\nonumber
\end{eqnarray}
where we have defined the operator $\vec{\pi}=\vec{p}-i\vec{\xi}$, whose vector $\vec{\xi}$ is given in such a way that its components are $-i\xi_{k}=-\frac{\sigma^{3}}{2\rho}\,\delta_{2k}$. Note that, the vector $\vec{\xi}$ in (\ref{5.3}) corresponds to the contribution from the spinorial connection \cite{bbs2} and gives rise to the relativistic correction term $\hat{H}_{\mathrm{spinorial}}=-\frac{g}{4m^{2}}\,\vec{\sigma}\cdot\left[\vec{E}_{\mathrm{eff}}\times\left(-i\vec{\xi}\right)\right]$.

From now on, let us consider the presence of a confining potential in order to study the influence of the Lorentz symmetry breaking scenario on a nonrelativistic Dirac neutral particle. We consider the nonrelativistic neutral particle confined to a two-dimensional quantum ring described by the Tan-Inkson model \cite{tan}. The Tan-Inkson potential \cite{tan} describes the confinement of a quantum particle to a two-dimensional quantum ring and, as a particular case, it describes the confinement of a quantum particle to a quantum dot. The Tan-Inkson model \cite{tan} consists in introducing the following scalar potential:
\begin{eqnarray}
V\left(\rho\right)=\frac{a_{1}}{\rho^{2}}+a_{2}\,\rho^{2}+V_{0},
\label{1.14}
\end{eqnarray}
where $V_{0}=2\sqrt{a_{1}a_{2}}$, $a_{1}$ and $a_{2}$ are control parameters. For $a_{1}=0$, we have that the Tan-Inkson model describes a two-dimensional quantum dot. For $a_{2}=0$, we have that the Tan-Inkson model describes a two-dimensional quantum anti-dot. Thereby, from the Lorentz symmetry violation background defined in Eqs. (\ref{5.0}) and (\ref{5.0a}), the Schr\"odinger-Pauli equation becomes
\begin{eqnarray}
i\frac{\partial\psi}{\partial t}&=&-\frac{1}{2m}\left[\frac{\partial^{2}}{\partial\rho^{2}}+\frac{1}{\rho}\frac{\partial}{\partial\rho}+\frac{1}{\rho^{2}}\frac{\partial^{2}}{\partial\varphi^{2}}+\frac{\partial^{2}}{\partial z^{2}}\right]\psi+\frac{1}{2m}\frac{i\sigma^{3}}{\rho^{2}}\frac{\partial\psi}{\partial\varphi}+\frac{1}{8m\rho^{2}}\psi+i\frac{gb_{0}\lambda B_{0}}{2m}\frac{\partial\psi}{\partial\varphi}\nonumber\\
&+&\frac{gb_{0}\lambda B_{0}}{4m}\,\sigma^{3}\,\psi+\frac{\left(gb_{0}\lambda B_{0}\right)^{2}}{8m}\,\rho^{2}\,\psi-\frac{gb_{0}\lambda^{2}}{4}\,\rho^{2}\,\psi-\frac{gb_{0}\lambda B_{0}}{2m}\,\sigma^{3}\,\psi-\frac{gb_{0}\lambda^{2}}{8m^{2}}\,\psi\label{5.4}\\
&+&i\,\frac{gb_{0}\lambda^{2}}{8m^{2}}\,\sigma^{3}\,\frac{\partial\psi}{\partial\varphi}-i\,\frac{gb_{0}\lambda^{2}\rho}{8m^{2}}\,\sigma^{2}\,\frac{\partial\psi}{\partial z}+\frac{gb_{0}\lambda^{2}}{16m^{2}}\,\psi\nonumber+\frac{a_{1}}{\rho^{2}}\,\psi+a_{2}\,\rho^{2}\,\psi+V_{0}\,\psi.
\end{eqnarray}

Note that $\psi$ is an eigenfunction of $\sigma^{3}$ in Eq. (\ref{5.4}), whose eigenvalues are $s=\pm1$. Thereby, we can write $\sigma^{3}\psi_{s}=\pm\psi_{s}=s\psi_{s}$. We can see that the operators $\hat{p}_{z}=-i\partial_{z}$ and $\hat{J}_{z}=-i\partial_{\varphi}$ \cite{schu} commute with the Hamiltonian of the right-hand side of (\ref{5.4}), therefore, we can write the solution of (\ref{5.4}) in terms of the eigenvalues of the operator $\hat{p}_{z}=-i\partial_{z}$ and the $z$-component of the total angular momentum $\hat{J}_{z}=-i\partial_{\varphi}$ \footnote{It has been shown in Ref. \cite{schu} that the $z$-component of the total angular momentum in cylindrical coordinates is given by $\hat{J}_{z}=-i\partial_{\varphi}$, where the eigenvalues are $\mu=l\pm\frac{1}{2}$.}: 
\begin{eqnarray}
\phi_{s}=e^{-i\mathcal{E}t}\,e^{i\left(l+\frac{1}{2}\right)\varphi}\,e^{ikz}\,R_{s}\left(\rho\right),
\label{5.5}
\end{eqnarray}
where $l=0,\pm1,\pm2,\ldots$ and $k$ is a constant. From now on, we consider $k=0$ in order to describe a planar system. Substituting the solution (\ref{5.5}) into the Schr\"odinger-Pauli equation (\ref{5.4}), we obtain the following radial equation:
\begin{eqnarray}
R_{s}''+\frac{1}{\rho}R_{s}'-\frac{\tau^{2}}{\rho^{2}}R_{s}-\frac{\left(gb_{0}\lambda B_{0}\delta\right)^{2}}{4}\,\rho^{2}\,R_{s}+\zeta_{s}\,R_{s}=0,
\label{5.6}
\end{eqnarray}
where we have defined in Eq. (\ref{5.6}) the parameters:
\begin{eqnarray}
\delta^{2}&=&1-\frac{2m}{gb_{0}B_{0}^{2}}+\frac{8m\,a_{2}}{\left(gb_{0}\lambda B_{0}\right)^{2}};\nonumber\\
\gamma_{s}&=&l+\frac{1}{2}\left(1-s\right)\nonumber\\
[-2mm]\label{5.7}\\[-2mm]
\tau&=&\gamma_{s}^{2}+2m\,a_{1}\nonumber\\
\zeta_{s}&=&2m\left(\mathcal{E}-V_{0}\right)+gb_{0}\lambda B_{0}\,\gamma_{s}+s\,gb_{0}\lambda B_{0}+s\,\frac{gb_{0}\lambda^{2}}{4m}\,\gamma_{s}+\frac{gb_{0}\lambda^{2}}{4m}.\nonumber
\end{eqnarray}

Let us make a change of variables given by $\xi=\frac{gb_{0}\lambda B_{0}\delta}{2}\,\rho^{2}$. In this way, the radial equation (\ref{5.6}) becomes
\begin{eqnarray}
\xi\,R_{s}''+R_{s}'-\frac{\tau^{2}}{4\xi}R_{s}-\frac{\xi}{4}\,R_{s}+\frac{\zeta_{s}}{2gb_{0}\lambda B_{0}\delta}\,R_{s}=0.
\label{5.8}
\end{eqnarray}

In order to obtain a regular solution at the origin, the solution of the second order differential equation (\ref{5.8}) is given by
\begin{eqnarray}
R_{s}\left(\xi\right)=e^{-\frac{\xi}{2}}\,\xi^{\frac{\left|\tau\right|}{2}}\,M_{s}\left(\xi\right).
\label{5.9}
\end{eqnarray}

Substituting (\ref{5.9}) into (\ref{5.8}), we obtain the following second-order differential equation:
\begin{eqnarray}
\xi\,M_{s}''+\left[\left|\tau\right|+1-\xi\right]M_{s}'+\left[\frac{\zeta_{s}}{2gb_{0}\lambda B_{0}\delta}-\frac{\left|\tau\right|}{2}-\frac{1}{2}\right]M_{s}=0.
\label{5.10}
\end{eqnarray}

Equation (\ref{5.10}) is the Kummer equation or the confluent hypergeometric function \cite{abra}. In order to obtain a solution for the equation (\ref{5.10}) regular at the origin, we consider only the Kummer function of first kind given by $M_{s}\left(\xi\right)=M\left(\frac{\left|\tau\right|}{2}+\frac{1}{2}-\frac{\zeta_{s}}{2gb_{0}\lambda B_{0}\delta},\,\left|\tau\right|+1,\,\xi\right)$ \cite{abra}. Thus, a normalized radial wave function can be obtained if we impose that the hypergeometric series becomes a polynomial of degree $n$. This makes the radial wave function to be finite everywhere \cite{landau}. Hence, a finite radial solution for Eq. (\ref{5.10}) can be achieved when the parameter $\frac{\left|\tau\right|}{2}+\frac{1}{2}-\frac{\zeta_{s}}{2gb_{0}\lambda B_{0}\delta}$ of the Kummer function is equal to a non-positive integer number, that is, when $\frac{\left|\tau\right|}{2}+\frac{1}{2}-\frac{\zeta_{s}}{2gb_{0}\lambda B_{0}\delta}=-n$ ($n=0,1,2,\ldots$). With this condition, the energy levels of the bound states are
\begin{eqnarray}
\mathcal{E}_{n,\,l,\,s}=\omega\delta\left[n+\frac{\left|\tau\right|}{2}+\frac{1}{2}\right]-\omega\,\frac{\gamma_{s}}{2}-s\,\frac{\omega}{2}-s\,\frac{gb_{0}\lambda^{2}}{8m^{2}}\,\gamma_{s}-\frac{gb_{0}\lambda^{2}}{8m^{2}}+V_{0},
\label{5.11}
\end{eqnarray}
where $\omega=\frac{gb_{0}\lambda B_{0}}{m}$ corresponds to the angular frequency.

Hence, the energy levels (\ref{5.11}) are characterized by different contributions that arise from the Lorentz symmetry breaking scenario defined in Eqs. (\ref{5.0}) and (\ref{5.0a}). The first contribution is obtained by analogy with the Landau system studied in Ref. \cite{lin2} confined to a two-dimensional quantum ring \cite{tan}. In this case, the energy levels associated with the analogue of the Landau system for a spin-$1/2$ neutral particle induced by Lorentz symmetry breaking effects confined to a two-dimensional quantum ring are:
\begin{eqnarray}
\mathcal{E}_{\mathrm{L}}=\omega\delta\left[n+\frac{\left|\tau\right|}{2}+\frac{1}{2}\right]-\omega\,\frac{\gamma_{s}}{2}.
\label{5.12}
\end{eqnarray}

It is worth mentioning that the parameter $\delta$ given in Eq. (\ref{5.7}) must be a positive number, that is, $\delta>0$. In this way, we have that both the radial equation (\ref{5.6}) and the energy levels (\ref{5.11}) do not have a imaginary part. The interesting fact behind the condition $\delta>0$ is that it allows us to establish an upper bound for the Lorentz symmetry breaking parameters. This occurs by imposing that $\frac{8m\,a_{2}}{\left(gb_{0}\lambda B_{0}\right)^{2}}-\frac{2m}{gb_{0}B_{0}^{2}}\,>\,0$. Thereby, let us consider $\lambda=10^{4}\,\mathrm{C}/\mathrm{m}^{3}$ and $a_{2}=2,222\cdot10^{-5}\,\frac{\mathrm{meV}}{\left(\mathrm{nm}\right)^{2}}$ as given in Ref. \cite{tan}, then, we obtain the following upper bound for the Lorentz symmetry breaking parameters
\begin{eqnarray}
g\,b_{0}\,<2,2\cdot10^{-6}\,\left(\mathrm{eV}\right)^{-3}.
\label{5.14}
\end{eqnarray}

On the other hand, the other contributions to the nonrelativistic energy levels (\ref{5.11}) correspond to the analogue of the Zeeman energy, $\mathcal{E}_{\mathrm{Z}}=-s\,\frac{\omega}{2}$, and the relativistic correction terms:
\begin{eqnarray}
\Delta\mathcal{E}=-s\,\frac{gb_{0}\lambda^{2}}{8m^{2}}\,\gamma_{s}-\frac{gb_{0}\lambda^{2}}{8m^{2}},
\label{5.13}
\end{eqnarray} 
which arise from the terms $H_{\mathrm{R}}$, $H_{\mathrm{D}}$ and $H_{\mathrm{spinorial}}$ induced by Lorentz symmetry breaking effects. Note that the first term of Eq. (\ref{5.13}) corresponds to an analogue of the Rashba energy which has been widely explored in studies of the Rashba coupling in condensed matter systems \cite{rasb,rasb2,rasel,rasel2,rasel3,rasel4,rasel5,rasmag}.

Hence, the effects of the violation of the Lorentz symmetry on the nonrelativistic Dirac neutral particle yield an analogue of the Landau system confined to a two-dimensional quantum ring. Moreover, we can see that the effects of the Lorentz symmetry violation background also yield a contribution to the energy levels (\ref{4.4}) that corresponds to an analogue of the Zeeman energy. The relativistic correction terms induced by Lorentz symmetry breaking effects yield new contributions to energy levels (\ref{5.12}) associated with the analogue of the Landau system confined to a two-dimensional quantum ring, where one of them is analogous to the Rashba effect \cite{rasb,rasb2,rasel,rasel2,rasel3,rasel4,rasel5,rasmag}.

\section{conclusions}

In this work, we have introduced a new coupling into the Dirac equation in order to study the arising of a Rashba-like coupling, a Zeeman-like term and a Darwin-like term in the Schr\"odinger-Pauli equation for a spin-$1/2$ neutral particle interacting with a field configuration of crossed electric and magnetic fields in a Lorentz symmetry violation background. We have shown, by considering a scenario of the Lorentz symmetry violation defined by a fixed time-like vector and the field configuration (\ref{5.0}), that an analogue of the Landau system can be achieved. Then, by including a confining potential which describes a two-dimensional quantum ring, we have obtained energy levels characterized by different contributions that arise from the Lorentz symmetry breaking effects. The first contribution to the energy levels is associated with the analogue of the Landau system for a spin-$1/2$ neutral particle confined to a two-dimensional quantum ring. Another contribution corresponds to the analogue of the Zeeman energy, while the remaining contributions correspond to the relativistic correction terms that gives rise to the analogue of the Rashba effect \cite{rasb,rasb2,rasel,rasel2,rasel3,rasel4,rasel5,rasmag}. Furthermore, we have shown that this possible scenario allows us to establish an upper bound for the parameters of the Lorentz symmetry breaking. Despite the difficulty of detecting Lorentz symmetry breaking effects and compare with any experimental data, this study allows us to perform an analytical analysis of the relativistic corrections terms, and opens new discussions of investigating the effects of the violation of the Lorentz symmetry at low energies.

\acknowledgments{We would like to thank CNPq (Conselho Nacional de Desenvolvimento Cient\'ifico e Tecnol\'ogico - Brazil) for financial support.}


\begin{thebibliography}{99}


\bibitem{extra3} V. A. Kostelecky and S. Samuel, Phys. Rev. D {\bf39}, 683 (1989);
									V. A. Kostelecky and S. Samuel, Phys. Rev. Lett. \textbf{63}, 224 (1989); 
								V. A. Kostelecky and S. Samuel, Phys. Rev. Lett. \textbf{66}, 1811 (1991); 
								V. A. Kostelecky and S. Samuel, Phys. Lett. B \textbf{381}, 89 (1996); 
								V. A. Kostelecky and R. Potting, Phys. Rev. D \textbf{51}, 3923 (1995).


\bibitem{extra1} D. Mattingly, Living Rev. Relativity {\bf8}, 5 (2005).

\bibitem{extra2} V. A. Kostelecky, {\it CPT and Lorentz Symmetry} (World Scientific, Singapore, 2011). 

\bibitem{col} D. Colladay and V. A. Kostelecky, Phys. Rev. D {\bf55}, 6760 (1997); 
							D. Colladay and V. A. Kostelecký, Phys. Rev. D {\bf58}, 116002 (1998);
							S. R. Coleman and S. L. Glashow, Phys. Rev. D {\bf59}, 116008 (1999);
							A. Kostelecky and M. Mewes, Phys. Rev. D {\bf80}, 015020 (2009);
							A. Kostelecky and M. Mewes, Phys. Rev. D {\bf85}, 096005 (2012).


\bibitem{lin3} L. R. Ribeiro. E. Passos and C. Furtado, J. Phys. G: Nucl. Part. Phys. {\bf39}, 105004 (2012). 

\bibitem{ea} E. O. Silva and F. M. Andrade, EPL {\bf101}, 51005 (2013).

\bibitem{ea2} F. M. Andrade, E. O. Silva, Phys. Lett. B {\bf719}, 467 (2013).

\bibitem{ea3} R. Casana, M. M. Ferreira Jr., E. Passos, F. E. P. dos Santos and E. O. Silva, Phys. Rev. D {\bf87}, 047701 (2013).

\bibitem{belich2} H. Belich, T. Costa-Soares, M. M. Ferreira Jr., J. A. Helay\"el-Neto and F. M. O. Moucherek, Phys. Rev. D \textbf{74}, 065009 (2006).

\bibitem{bb2} K. Bakke and H. Belich, Eur. Phys. J. Plus {\bf127}, 102 (2012).

\bibitem{bb3} K. Bakke and H. Belich, Ann. Phys. (NY) {\bf333}, 272 (2013).

\bibitem{belich} H. Belich, T. Costa-Soares, M. M. Ferreira Jr. and J. A. Helay\"el-Neto, Eur. Phys. J. C \textbf{41}, 421 (2005).

\bibitem{belich1} H. Belich, T. Costa-Soares, M. M. Ferreira Jr., J. A. Helay\"el-Neto, M. T. D. Orlando, Phys. Lett. B \textbf{639}, 675 (2006).

\bibitem{belich3} H. Belich, L. P. Collato, T. Costa-Soares, J. A. Helay\"el-Neto and M. T. D. Orlando, Eur. Phys. J. C \textbf{62}, 425 (2009).

\bibitem{bbs2} K. Bakke, H. Belich and E. O. Silva, Ann. Phys. (Berlin) {\bf523}, 910 (2011).

\bibitem{bb} K. Bakke and H. Belich, J. Phys. G: Nucl. Part. Phys. {\bf39}, 085001 (2012).

\bibitem{brito} M. A. Anacleto, F. A. Brito and E. Passos, Phys. Rev. D {\bf86}, 125015 (2012).

\bibitem{bb4} K. Bakke and H. Belich, J. Phys. G: Nucl. Part. Phys. {\bf40}, 065002 (2013).

\bibitem{rasb} E. I. Rashba, Fiz. Tverd. Tela (Leningrad) {\bf2}, 1224 (1960) [Sov. Phys. Solid State {\bf2}, 1109 (1960)];
								Yu. A. Bychkov and E. I. Rashba, J. Phys. C {\bf17}, 6039 (1984).		

\bibitem{rasb2} E. Tsitsishvili, G. S. Lozano and A. O. Gogolin, Phys. Rev. B {\bf70}, 115316 (2004).

\bibitem{rasel} E. I. Rashba, Physica E {\bf20}, 189 (2004).

\bibitem{rasel2} A. V. Moroz and C. H. W. Barnes, Phys. Rev. B {\bf61}, R2464 (2000).

\bibitem{rasel3} O. Voskoboynikov, C. P. Lee and O. Tretyak, Phys. Rev. B {\bf63}, 165306 (2001).

\bibitem{rasel4} S.-Q. Shen, Pys. Rev. Lett. {\bf95}, 187203 (2005).

\bibitem{rasel5} S. Bellucci and P. Onorato, Phys. Rev. B {\bf68}, 245322 (2003).

\bibitem{rasmag} J. Schliemann, Phys. Rev. B {\bf77}, 125303 (2008).

\bibitem{tan} W.-C. Tan and J. C. Inkson, Semicond. Sci. Technol. {\bf11}, 1635 (1996); 
							W.-C. Tan and J. C. Inkson, Phys. Rev. B {\bf53}, 6947 (1996);  
							W.-C. Tan and J. C. Inkson, Phys. Rev. B {\bf60}, 5626 (1999).


\bibitem{fw} L. L. Foldy and S. A. Wouthuysen, Phys. Rev. {\bf78}, 29 (1950).

\bibitem{greiner} W. Greiner, \textit{Relativistic Quantum Mechanics: Wave Equations, 3rd Edition} (Springer, Berlin, 2000).

\bibitem{bb5} K. Bakke and H. Belich, Ann. Phys. (Berlin) {\bf526}, 187 (2013).


\bibitem{lin2} C. Furtado, J. R. Nascimento and L. R. Ribeiro, Phys. Lett. A {\bf358}, 336 (2006).

\bibitem{weinberg} S. Weinberg, {\it Gravitation and Cosmology: Principles and Applications of the General Theory of Relativity} (IE-Wiley, New York, 1972).

\bibitem{bd}  N. D. Birrell and P. C. W. Davies, \textit{Quantum Fields in Curved Space}, (Cambridge University Press, Cambridge, UK, 1982).

\bibitem{schu} P. Schl\"uter, K.-H. Wietschorke and W. Greiner, J. Phys. A {\bf16}, 1999 (1983).

\bibitem{b4} K. Bakke, Ann. Phys. (Berlin) {\bf523}, 762 (2011).

\bibitem{abra} M. Abramowitz and I. A. Stegum, \textit{Handbook of mathematical functions} (Dover Publications Inc., New York, 1965).

\bibitem{landau} L. D. Landau and E. M. Lifshitz, \textit{Quantum Mechanics, the nonrelativistic theory, 3rd Ed.} (Pergamon, Oxford, 1977).










\end{thebibliography}
\end{document}